\documentclass[12pt,preprint]{aastex}
\usepackage{emulateapj5}

\shortauthors{T.Hosokawa \& S.Inutsuka}
\shorttitle{A Starburst Mechanism}

\begin{document}

\title{Dynamical Expansion of Ionization and Dissociation Front around
a Massive Star : A Starburst Mechanism}
\author{Takashi Hosokawa\altaffilmark{1} and 
     Shu-ichiro Inutsuka\altaffilmark{2}}

\altaffiltext{1}{Division of Theoretical Astrophysics, 
National Astronomical Observatory, Mitaka, Tokyo 181-8588, Japan ; 
hosokawa@th.nao.kyoto-u.ac.jp}
\altaffiltext{2}{Department of Physics, Kyoto University, Kyoto 606-8502 ;
inutsuka@tap.scphys.kyoto-u.ac.jp} 

\begin{abstract}
We quantitatively examine the significance of star formation 
triggered in the swept-up shell around an expanding H~II region.
If the swept-up molecular gas is sufficiently massive,
new OB stars massive enough to repeat the triggering process
will form in the shell. We determine the lower limit ($M_{\rm thr}$) 
for the mass of the star that sweeps up the molecular gas, 
where at least one new star with mass $M_* > M_{\rm thr}$ 
forms after the shell fragmentation.
To calculate the threshold stellar mass, $M_{\rm thr}$, 
 we examine how massive molecular shells can form around various 
 central stars, by performing detailed numerical radiation 
 hydrodynamics calculations.
The mass of the photodissociated gas is generally larger than
 the mass of the photoionized gas.
However, the swept-up molecular mass exceeds the photodissociated
 mass with a higher-mass star of $M_* \gtrsim 20~M_\odot$.
The accumulated molecular mass generally increases with the stellar
 mass, and amounts to $10^{4-5}~M_\odot$ for $M_* \gtrsim 20~M_\odot$
 with an ambient density of $n \sim 10^2~{\rm cm}^{-3}$.
The threshold stellar mass is $M_{\rm thr} \sim 18~M_\odot$ 
 with the star-formation efficiency of $\epsilon \sim 0.1$ 
 and $n \sim 10^2~{\rm cm}^{-3}$.
We examine the generality of this mode of run-away triggering for 
 different sets of parameters, and found that 
 $M_{\rm thr} \sim 15-20~M_\odot$ in various situations.  
If the ambient density is too high or the star-formation efficiency 
 is too low, the triggering is not run-away, but a single event. 
\end{abstract}
\keywords{ Circumstellar matter -- H~II regions -- ISM: molecules 
           -- STARS : formation}

\section{Introduction}

In galaxies, most of the stars are born in localized
active star-forming regions or starburst regions.
Clustered star formation is the dominant mode in these
regions \citep[e.g.,][]{Ev99, LL03}, and several star clusters, 
including OB stars, form over a few Myrs from the parental giant 
molecular cloud \citep{WM97}. The newly born massive stars 
promptly emit the UV ($h \nu > 13.6~{\rm eV}$) and far-UV
(FUV ; $11~{\rm eV} \lesssim h \nu < 13.6~{\rm eV}$) 
radiation, and the radiative feedback by these photons occurs 
much earlier than the supernova explosions.
The UV and FUV radiation from massive stars has
two competing effects on the parental molecular cloud.
One is the {\it negative feedback} effect on the star formation activity.
The parental molecular cloud is ionized and heated up by the UV
radiation, and the nearby star formation is quenched
\citep[e.g.,][]{Wt79, FST94, WM97, Mz02}.
\citet{DFS98} have shown that the photodissociation by FUV photons
is more significant than the photoionization by UV photons.
The other is the {\it positive feedback} effect. 
The next star formation is triggered in the compressed dense layer 
around the H~II region 
\citep[collect and collapse scenario ; e.g.,][]{EL77, Elm89}. 
Recent observations have provided dramatic snapshots of 
 this triggering process \citep[e.g.,][]{Dv03, Dv05, Zv05}.
\citet{KHL05} have observed details of the starburst
 regions, and suggested that the propagating star formation is
important on the scale of the clouds. 
The character of star formation depends on which feedback effect 
 dominates the other in the cloud.

In our previous papers, \citet{HI05,HI052} 
(hereafter Papers I and II), we have studied the time
evolution of the H~II region, photodissociation region (PDR),
and the swept-up shell, performing detailed numerical calculations.
We have shown that the molecular gas is accumulated around the 
H~II region, and that the gravitational fragmentation of the shell is 
expected in many cases with the homogeneous molecular ambient medium. 
This suggests that the expanding H~II region is an efficient
trigger of the star formation in the cloud.
In this letter, we examine the net feedback effect of the expanding 
H~II region and PDR around a massive star.

\section{A Starburst Mechanism}

\subsection{Triggering Threshold Condition}

First, we focus on the positive feedback process for triggering
 the star formation in the shell.
Let us consider a situation where a dense molecular
 shell, whose mass is $M_{\rm sh,m}$, forms around a massive 
 star, and the subsequent star formation is triggered after
fragmentation of the shell. 
The total number of newly born stars is calculated as  
\begin{equation}
N_* = \frac{M_{\rm sh, m} \epsilon}{M_{\rm av}} ,
\label{eq:nstar}
\end{equation}
 where $\epsilon$ is the star formation efficiency (SFE) 
 for the swept-up gas and 
 $M_{\rm av}$ is the average stellar mass, which is defined as,
\begin{equation}
M_{\rm av} \equiv \int^\infty_0
                  \phi(M) M~dM  \bigg/ 
                  \int^\infty_0 \phi(M)~dM ,
\end{equation}
where $\phi(M)$ is the initial mass function (IMF).
Below, we use the IMF by \citet{MS79}.
The calculated value of $M_{\rm av}$ is about $0.6~M_\odot$.
The mass of the swept-up molecular gas, $M_{\rm sh, m}$, 
 generally depends on the ambient number density and 
 mass of the central star (see Paper II). 
If $M_{\rm sh,m}$ is sufficiently large, another massive star 
as well as lower-mass stars will form in the shell. 
Another H~II region expands around this newly born massive star, 
 and the triggering process will repeat. 
Furthermore, if the number of newly-born stars is larger than that 
 of the previous generation, this process causes run-away
 triggering or a burst of star formation.

In order to examine the efficiency of this mode of triggering, 
we presume that the run-away triggering begins with
a star more massive than the {\it threshold stellar mass},
$M_{\rm thr}$.
A star of $M_{\rm thr}$ produces a shell that barely generates
just one star more massive than $M_{\rm thr}$.
If the triggering cycle begins with a star of $M_* > M_{\rm thr}$,
multiple stars of $M_* > M_{\rm thr}$ form as ``second
generation'' stars. The number of newborn stars increases as the 
triggering cycle advances, and the burst of star formation occurs.
We represent the condition for the run-away triggering with simple
 equations. 
The number of $M_* > M_{\rm thr}$ stars born in the shell is,
\begin{equation}
N_{\rm thr} = f_{\rm thr} N_* ,
\label{eq:nthr}
\end{equation}
where $f_{\rm thr}$ is the number ratio of $M_* > M_{\rm thr}$
stars calculated as,
\begin{equation}
f_{\rm thr} \equiv \int^\infty_{M_{\rm thr}}
                  \phi(M)~dM  \bigg/ 
                  \int^\infty_0
                  \phi(M)~dM .
\label{eq:cond}
\end{equation}
We can determine $M_{\rm thr}$ from the 
{\it triggering threshold condition}; $N_{\rm thr} = 1$.
Combining this condition and equations (\ref{eq:nstar}) and
(\ref{eq:nthr}), we obtain
\begin{equation}
   M_{\rm sh, m} = \frac{M_{\rm av}}{\epsilon f_{\rm thr} } .
\label{eq:fthr}
\end{equation}
Evaluating how $M_{\rm sh, m}$ depends on the stellar mass,
we can calculate $M_{\rm thr}$ for a given ambient number density 
and SFE by equation (\ref{eq:fthr}). If the molecular mass of the 
shell is smaller than $M_{\rm av} / \epsilon f_{\rm thr}$, 
 triggering is not run-away, but only a single event. 
We examine $M_{\rm sh,m}$ with various stellar masses and 
 ambient number densities based on numerical calculations.

\subsection{Mass of the Molecular Shell v.s. Mass of the Central Star}
\label{ssec:shmass}

In order to evaluate the threshold stellar mass, $M_{\rm thr}$
 for the burst of star formation, we examine how the shell mass,
 $M_{\rm sh,m}$ changes with the mass of the massive star 
 and the ambient molecular gas density. 
Even if the shell forms around the H~II region, the accumulated
gas is exposed to FUV radiation from the central star.
Unless the FUV radiation is blocked, almost no cold
molecular gas accumulates in the shell.   
Therefore, we should carefully estimate the mass of the 
 cold molecular gas in the shell, which is available for 
 the next star-formation episode. 
For this purpose, we have calculated the 
time evolution of the H~II region, PDR, and the swept-up shell around 
various central stars in the homogeneous molecular medium 
(see Paper II for the details).
In this subsection, we focus on the fiducial case with 
$n_{\rm H} = 100~{\rm cm}^{-3}$ ($n_{\rm H_2} = 50~{\rm cm}^{-3}$).
If the FUV radiation is efficiently shielded in the shell, 
the accumulated molecular mass, $M_{\rm sh,m}$ increases 
as the H~II region expands. However, $M_{\rm sh,m}$ is actually 
 limited by the following two factors: 
\begin{center}
\epsfxsize=9cm
\epsfbox{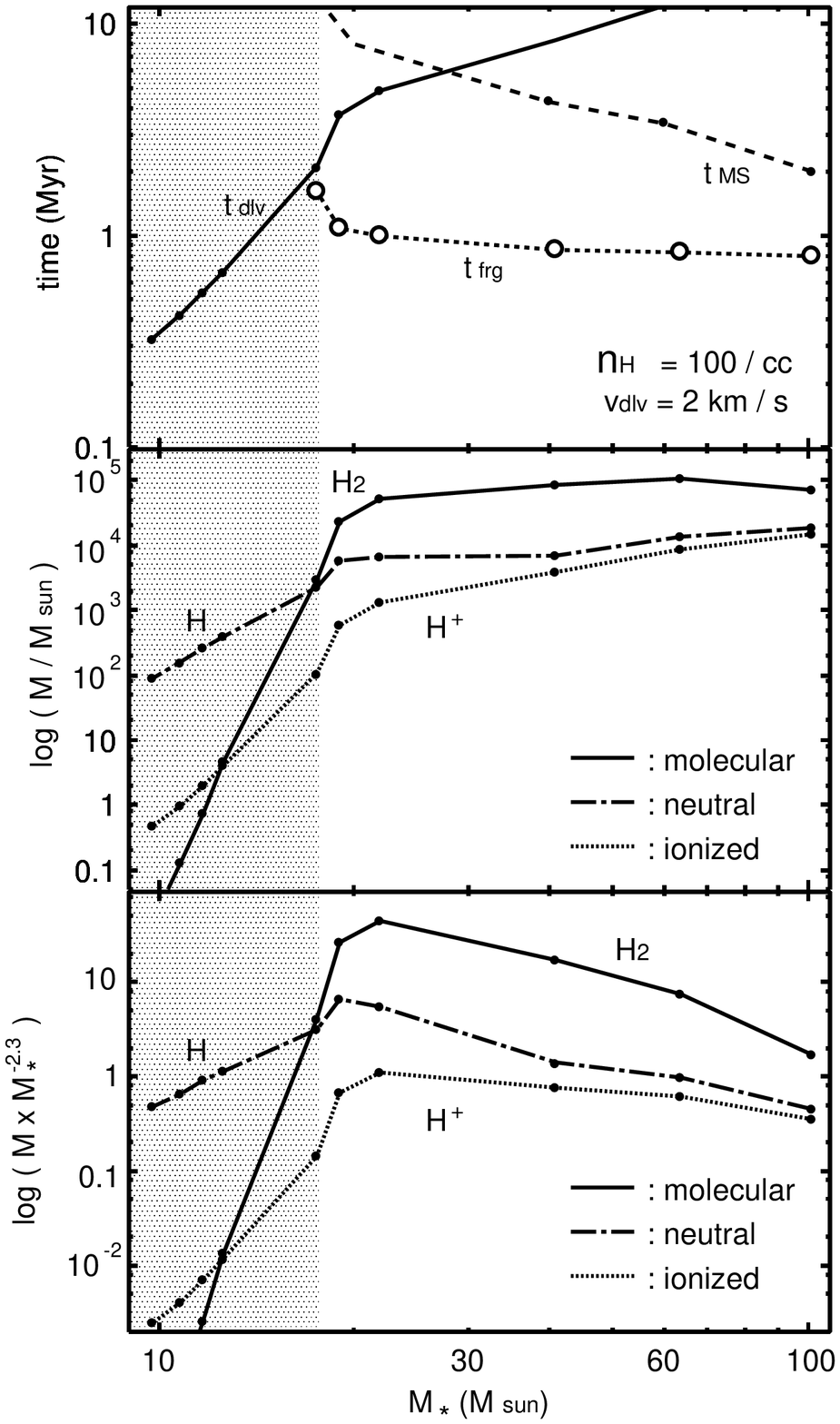}
\figcaption{{\it Top panel : } 
Stellar mass v.s. the termination time of the shell
in the ambient medium of $n_{\rm H} = 100~{\rm cm}^{-3}$ with the 
dissolving velocity, $v_{\rm dlv} = 2~{\rm km/s}$.
The solid and broken lines represent the dissolving time
of the shell, $t_{\rm dlv}$ and the main-sequence lifetime
of the central star, $t_{\rm MS}$ respectively. 
The termination time of the shell is the smaller one
between $t_{\rm dlv}$ and $t_{\rm MS}$. The dotted line means the time 
when the gravitationally unstable region appears in the shell. 
The dots and open circles indicate the calculated values in each model.
{\it Middle panel : } The mass of molecular hydrogen 
accumulated in the shell (solid line), neutral hydrogen
in the PDR (dot-dashed line), and ionized hydrogen in the
H~II region (dotted line) for various central stars at the 
termination time. {\it Bottom panel : } The relative fraction of three 
masses presented in the middle panel. We plot 
$M \times M_*^{-2.3} \propto M \phi(M_*)$, where $\phi(M_*)$ is the 
IMF given by \citet{MS79}. In each panel, a positive feedback
effect is expected in the unshaded mass range, while negative
is expected in the shaded range.}
\label{fig:fdbk_50_2km}
\end{center}
First, the expanding shell finally 
 dissolves due to the turbulent motion of the ambient medium. 
Ubiquitous supersonic turbulent motions in molecular clouds 
 have been recognized by the observed broad emission line width.   
The corresponding velocity dispersion increases with the 
 length scale as $\sigma \propto l^{1/2}$, and 
 $\sigma \sim 2~{\rm km/s}$ at $l \sim 10~{\rm pc}$. 
We define the dissolving velocity, 
 $v_{\rm dlv} \sim {\rm several} \times {\rm km/s}$, and 
 the shell dissolves at the time, $t_{\rm dlv}$, when
 the shell velocity becomes less than $v_{\rm dlv}$. 
Second, OB stars evolve to the Wolf-Rayet phase after 
 the main-sequence phase, and finally explode as supernovae. 
Although the Wolf-Rayet winds and supernova explosions
 must cause the dynamical expansion of bubbles, we do not
 consider the evolution after the main-sequence lifetime, $t_{\rm MS}$.
We evaluate the mass of the molecular gas in the shell
at the termination time of the shell, $\min(t_{\rm dlv}, t_{\rm MS})$.
The top panel of Figure 1 
presents $t_{\rm dlv}$ and $t_{\rm MS}$ with various central
stars for $n_{\rm H} = 100~{\rm cm}^{-3}$ and 
$v_{\rm dlv} = 2~{\rm km/s}$. As this panel shows, the dissolving 
time is shorter for the lower-mass star. This is because the dynamical 
time, $t_{\rm dyn}$ is shorter for the less luminous central star as 
$t_{\rm dyn} \equiv R_{\rm st}/C_{\rm II} \propto S_{\rm UV}^{1/3}$ 
(Paper II), where $S_{\rm UV}$ is the UV photon number luminosity.
With shorter $t_{\rm dyn}$, the expansion quickly decelerates
and terminates earlier. The main-sequence 
lifetime, $t_{\rm MS}$ is shorter for the higher-mass 
star, and the expansion is limited by $t_{\rm MS}$ for 
$M_* \gtrsim 30~M_\odot$.

The middle panel of Figure 1 presents the mass
 of the swept-up molecular gas at the termination time.
This panel shows that $M_{\rm sh,m}$ increases with the stellar
mass, which verifies the triggering
threshold condition.
The molecular mass amounts to $M_{\rm sh,m} \sim 10^{4-5}~M_\odot$
 for the massive central stars of $M_* \gtrsim 20~M_\odot$.
In these cases, the H~II region expands to $10-20$~pc by the termination
 of the shell. 
With the central star of $M_* \lesssim 20~M_\odot$,
 the accumulated molecular mass is significantly small. 
The shell dissolves before the accumulation of the molecular gas with 
 the lower-mass star. 
To clarify the dominant feedback effect, 
 we also plot the masses of the photodissociated and 
 photoionized gas around the star at the same time.
With the higher-mass stars of $M_* \gtrsim 20~M_\odot$, the
accumulated molecular mass is much larger than both the ionized and 
photodissociated masses. Therefore, such massive stars give have a
positive feedback effect on the clouds.
With the lower-mass stars of $M_* \lesssim 20~M_\odot$,
on the other hand, the photodissociated mass is much larger than 
the ionized and accumulated molecular masses. 
In these cases, the net feedback effect is negative, and the
photodissociation by FUV photons is the dominant negative process, 
as argued by \citet{DFS98}. 
The bottom panel of Figure 1 presents the masses
multiplied by the population of exciting stars. 
When the higher-mass stars of $M_* \gtrsim 20~M_\odot$ are formed in
the cloud, the positive feedback effect can dominate, despite
a large number of lower-mass stars. 
Whether the net feedback effect is negative or positive depends on
the mass of the most massive star in the cloud; positive for 
$M_* \gtrsim 20~M_\odot$, and vice versa in the current fiducial
case.

\subsection{Threshold Stellar Mass for Starburst}

Using the calculated $M_{\rm sh,m}$ for various central stars,
we determine the threshold stellar mass, $M_{\rm thr}$ for the
burst of star formation. For example, in the fiducial case with
$n_{\rm H} = 100~{\rm cm}^{-3}$ and $v_{\rm dlv} = 2$~km/s,
the threshold mass calculated by equation (\ref{eq:cond}) is
$M_{\rm thr} \sim 18~M_\odot$ with $\epsilon = 0.1$.
This means that only a star more massive than $18~M_\odot$ can
generate a shell massive enough so as to produce at 
least one star as massive as the original star.
\vspace{-1cm}
\begin{center}
\epsfxsize=9cm
\epsfbox{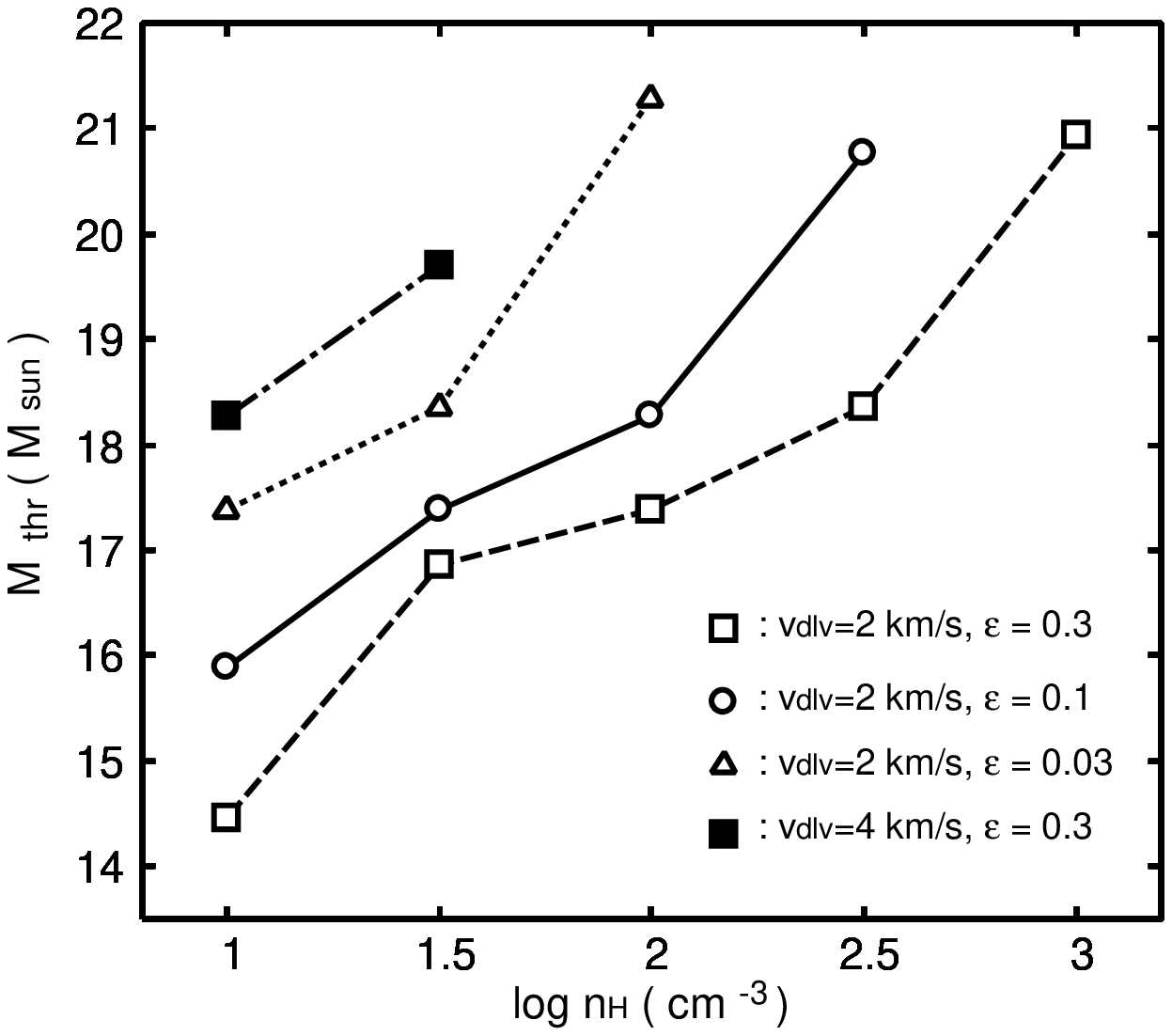}
\figcaption{Threshold stellar masses, $M_{\rm thr}$ for the run-away
triggering with different parameters. 
The horizontal axis is the ambient number density, and the different
symbols denote different sets of dissolving velocity, 
$v_{\rm dlv}$ and star formation efficiency, $\epsilon$.
Some symbols disappear at higher ambient densities, which means
that the triggering threshold condition (eq.(\ref{eq:fthr})) is
not satisfied with the corresponding set of parameters. 
In these cases, triggering is not run-away, but a single event.
}
\label{fig:m_thr}
\end{center}
\vspace{0mm}
Note that the threshold mass is comparable to the stellar mass 
over which the character of feedback changes from negative to positive. 
In this case with $M_* \gtrsim 20~M_\odot$, therefore, 
the net feedback effect is positive, and strong enough
to cause the run-away triggering. 

We have examined the possibility of run-away triggering
 by calculating $M_{\rm thr}$ with different parameters (Fig.2). 
For example, $M_{\rm thr}$ increases with decreasing SFE.  
This is because the more massive molecular shell is needed to 
 compensate for the lower SFE. With $\epsilon = 0.03$ in the 
fiducial case, we obtain $M_{\rm thr} \sim 21~M_\odot$. 
If the SFE is as low as $\epsilon \lesssim 0.01$, 
 however, equation (\ref{eq:cond}) 
 is not satisfied with any $M_{\rm thr}$. 
This is because the increase in $M_{\rm sh,m}$ grows saturated 
 at about $\sim 10^5~M_\odot$ as the mass of the central star 
increases (see Fig.1). Although the net feedback effect is positive with 
 $M_* \gtrsim 20~M_\odot$, triggering is not run-away, but
 only a single event.

The threshold stellar mass also increases with the higher
ambient density. In the denser ambient medium, the size of the 
H~II region shrinks as $R_{\rm st} \propto n_{\rm H}^{-2/3}$.
The total swept-up mass at the termination time
also becomes smaller following the scaling relation,
$M_{\rm sh} \propto n_{\rm H}^{-1}$,
with the same central star. With the higher ambient density, 
therefore, $M_{\rm thr}$ increases to 
produce the massive shell which enables the run-away triggering.
Since the increase in $M_{\rm sh,m}$ becomes saturated for the 
higher-mass stars, the run-away triggering becomes impossible unless 
 the SFE is very high in the denser ambient medium.
With $n_{\rm H} = 10^3~{\rm cm}^{-3}$ and 
 $v_{\rm dlv} = 2~{\rm km/s}$, for example, an SFE as high as 
 $\epsilon \sim 0.3$ is needed for the run-away triggering.

Finally, the higher dissolving velocity reduces the significance
of triggering. The expanding shell dissolves earlier with higher 
$v_{\rm dlv}$, and the H~II region sweeps up only the smaller region.
Furthermore, the molecular fraction in the shell 
is preferentially reduced with the higher $v_{\rm dlv}$.
During the dynamical expansion of the H~II region, 
the column density of the shell, $N_{\rm sh}$ increases as
$N_{\rm sh} \sim n_{\rm H}~R_{\rm IF}(t)/3$.
Therefore, the smaller H~II region has a shell with lower 
column density, though the incident FUV radiation is strong owing to the 
poor geometrical dilution. Since the FUV radiation shielding is 
inefficient, it is hard for molecules to accumulate in the shell.
With $v_{\rm dlv} = 4~{\rm km/s}$ and $\epsilon = 0.3$, for example, 
the run-away triggering is possible only with
the low ambient density of $n_{\rm H} \lesssim 10^{1.5}~{\rm cm}^{-3}$.

\section{Discussions and Conclusions}

In this letter, we suggest a possible mode of self-propagating
 star formation in molecular clouds, which should be examined in 
 further studies. 
Although we have adopted the homogeneous ambient density, for example,
real clouds show clumpy structure with a turbulent
velocity field. Recently, \citet{Ml06} have calculated the dynamical
expansion of the H~II region in turbulent molecular clouds.
They have shown that the clumpy medium leads to an irregular
ionization front. 
The radial position of the ionization front approximately agrees
with that in the homogeneous medium
on average with the mean density, 
but significantly depends on the radial direction.    
More theoretical studies and multi-dimensional numerical simulations 
of the deformation and fragmentation processes of
 the swept-up shell are needed. 
Observational estimation of the SFE in the shell might be 
 interesting and useful for the present model. 

The stellar wind from the massive star is another omission in
our evaluation. The expanding wind-driven bubble can modify
the dynamics of the ISM around massive stars.
When the bubble pressure is much higher than the ambient pressure,
the time evolution of the bubble size and pressure is given by,
\begin{equation}
R_{\rm b}(t) =  \left( \frac{125}{154 \pi} \right)^{1/5}
                L_{\rm w}^{1/5} \rho_0^{-1/5} t^{3/5}, 
\label{eq:rb}
\end{equation}
\begin{equation}
P_{\rm b}(t) = \frac{7}{(3850 \pi)^{2/5}} 
               L_{\rm w}^{2/5} \rho_0^{3/5} t^{-4/5} ,
\label{eq:pb}
\end{equation}
where $L_{\rm w}$ is the wind mechanical luminosity \citep{Wv77}.
Here, we evaluate
the significance of the wind-driven bubble by comparing
the ratio between the bubble pressure and H~II pressure 
($P_{\rm II} = \rho_0 C_{\rm II}^2$)
at $R_{\rm b} = R_{\rm st}$.
With equations (\ref{eq:rb}) and (\ref{eq:pb}), 
\begin{equation}
f_{\rm P} \equiv 
 \left. \frac{P_{\rm b}}{P_{\rm II}} \right|_{R_{\rm b} = R_{\rm st}}
 = \frac{10}{11} 
   \left( \frac{77}{250} \right)^{1/3}
   f_{\rm L}^{2/3}
 \sim \left( \frac{f_{\rm L}}{2} \right)^{2/3}, 
\end{equation}
where we define $f_{\rm L} \equiv L_{\rm w}/L_{\rm st}$ and 
$L_{\rm st} \equiv 4 \pi R_{\rm st}^2 \rho_0 C_{\rm II}^3$
\citep{MK84}. 
If $f_{\rm P} \lesssim 1$, the bubble is confined by the ambient H~II
pressure before reaching the initial Str\"omgren radius.
The numerical value of $f_{\rm L}$ can be estimated by 
$f_{\rm L} = 0.8 (S_{\rm UV}/10^{49} {\rm s}^{-1})^{5/6} 
n_{\rm H}^{1/3}$ \citep{Ab82}.
With the number density of $n_{\rm H} = 100~{\rm cm}^{-3}$,   
$f_{\rm P} \sim 0.4$ for a $20~M_\odot$ star and
$f_{\rm P} \sim 1.5$ for a $50~M_\odot$ star.
Therefore, the stellar wind only slightly increases the initial
pressure of the H~II region with the star of $M_* \lesssim 50~M_\odot$.
The effect of the wind-driven bubble becomes significant with
a very massive star ($\sim 100~M_\odot$), highly clustered
massive stars, or in higher density media.
When a wind-driven bubble dominates the dynamics, the bubble
sweeps up the larger region at higher velocity than
the expansion only due to the H~II region overpressure. 
This situation may even promote the positive feedback effect,
and should be separately studied in detail. 

Finally, we summarize the results. 
First, 
 we have formulated the conditions for the positive feedback process 
 by which the triggered star formation continues in the swept-up shell.
The threshold stellar mass, $M_{\rm thr}$, is defined as 
 the mass of the central star that sweeps up the molecular gas 
 in the shell, where at least one new star with mass 
 $M_* = M_{\rm thr}$ forms after the shell fragmentation.
In order to evaluate $M_{\rm thr}$, we have calculated how massive
 molecular shells can form around various central stars.
In the fiducial case with $n_{\rm H} = 100~{\rm cm}^{-3}$ and
 $v_{\rm dlv} = 2~{\rm km/s}$, the swept-up molecular mass increases
 with the stellar mass, 
 and amounts to $M_{\rm sh,m} \sim 10^{4-5}~M_\odot$ 
 with the massive star of $M_* \gtrsim 20~M_\odot$.
We have also calculated the photoionized and photodissociated masses
 to clarify the net feedback effect.
The negative effect of photodissociation is important with
 the lower-mass star of $M_* \lesssim 20~M_\odot$, 
 but is dominated by the positive effect with the higher-mass star.
We have calculated the threshold mass as $M_{\rm thr} \sim 18~M_\odot$
 in the fiducial case with $\epsilon = 0.1$.
We have also calculated $M_{\rm thr}$ with different parameters,
 and examined the generality of run-away triggering.
We have found $M_{\rm thr} \sim 15-20~M_\odot$ in various
 situations. The triggering process is only a single event with 
the higher ambient density, higher dissolving velocity, or lower SFE.

{\acknowledgements 
We thank a anonymous referee for fruitful comments that
improved the presentation of the paper. 
SI is supported by the Grant-in-Aid  (15740118, 16077202)
from the Ministry of Education, Culture, Sports, Science, and
Technology (MEXT) of Japan.}

\end{document}